\documentstyle[preprint,aps]{revtex}
\begin{document}
\title{
Quantum Critical Scaling in a Moderately Doped Antiferromagnet}
\author{\large Alexander Sokol}
\address{\rm
Department of Physics and Materials Research Laboratory, \\
University of Illinois at Urbana-Champaign, Urbana, IL 61801-3080 \\
and L.D.\ Landau Institute, Moscow, Russia}
\author{\large Rodney L. Glenister and Rajiv R.P. Singh}
\address{\rm
Department of Physics, University of California, Davis, CA 95616}
\maketitle
\begin{abstract}
Using high temperature expansions for the equal time correlator
$S(q)$ and static susceptibility $\chi(q)$
for the t-J model, we present evidence for quantum critical (QC),
$z\!=\!1$,
behavior at intermediate temperatures
in a broad range of $t/J$ ratio, doping, and temperatures.
We find that the dynamical susceptibility is very close to the
universal scaling function computable for the asymptotic
QC regime, and that the dominant energy scale is temperature.
Our results are in excellent agreement
with measurements of the spin-echo decay rate,
$1/T_{\rm 2G}$, in La$_2$CuO$_4$, and provide qualitative
understanding of both $1/T_1$ and $1/T_{\rm 2G}$ nuclear
relaxation rates in doped cuprates.
\end{abstract}
\pacs{}
\narrowtext
\rm
Recent interest in the doped antiferromagnets is related
to the problem of high temperature superconductivity.
While the magnetic behavior of the parent insulating compound,
La$_2$CuO$_4$, can be described in terms of the S=1/2 Heisenberg
model with the dominant interaction being
the in-plane nearest neighbor
exchange coupling $J\!\simeq\!1500$K, current understanding
of the doped materials is far from complete.
The consensus on the details of the microscopic model which would
quantitatively describe the magnetic properties over the entire
doping range from the insulator to the fully doped compounds
has yet to be reached.

Recently, significant progress has been made in
understanding the low energy spin dynamics of these systems
from a scaling and renormalization group point of view.
As shown by Chakravarty, Halperin, and Nelson \cite{CHN},
the spin dynamics of an insulator, such as La$_2$CuO$_4$,
is well described by the quantum nonlinear sigma
(QNL$\sigma$) model.
In case when the average sublattice magnetization
is present at $T=0$, the low temperature
renormalized classical (RC) phase is characterized by an
exponentially increasing antiferromagnetic correlation length,
$\xi/a\sim\exp(2\pi\rho_s/T)$,
where $a$ is the lattice constant and $\rho_s$ spin stiffness
(below we assume the units where $\hbar\!=\!k_B\!=\!a\!=1$).
Beyond the critical point, i.e.,
when the ground state does not possess Neel order,
the quantum disordered (QD) phase has a
finite $\xi$ at T=0.

Another aspect of the phase diagram of Ref.\cite{CHN}, which did
not attract much attention until the recent work of Sachdev and Ye
\cite{Sachdev:Ye}, and Chubukov and Sachdev \cite{Chubukov:Sachdev},
is the quantum critical (QC) region, where
in the leading order $\xi\!\sim\!1/T$ \cite{CHN}.
Only at the critical point, i.e. at the boundary between
the zero temperature Neel and disordered phases, QC behavior holds
down to $T\!=\!0$. In this case, $\rho_s$
vanishes, so that the model does not possess any energy
scale, which thus is set by the temperature.
For doped antiferromagnets, the possibility of $z$=1 criticality has
been first pointed out in Ref.\cite{Sachdev:Ye}.

Although for small temperatures a fine-tuning of the model parameters
to the critical point is necessary
to have the QC phase with $\xi\!\sim\!c/T$, the range of parameters
where it exists rapidly
widens as the temperature increases ($c$ is the spin wave velocity).
While in the continuum QNL$\sigma$ model the quantum critical behavior
persists for arbitrarily high temperature, on the lattice the range
of its applicability is necessarily limited form above, i.e.\
quantum critical
regime is intermediate between low and high temperatures.
It has been argued in Ref.\cite{Chubukov:Sachdev}
that the region of its applicability in the 2D S=1/2 Heisenberg model
exists around $T\sim 0.5J$ and that
small doping should lead to a decrease
in $\rho_s$, thus extending the temperature range of
QC behavior.

Recently, it has been shown by Pines and one of us (A.S.)
\cite{Sokol:Pines} using purely scaling considerations, that
the experimental data of \cite{ImaiT1,ImaiT2,TakigawaT1,TakigawaT2}
on the nuclear magnetic
relaxation rates $1/T_1$ and $1/T_{\rm 2G}$ in the
superconducting cuprates imply a QC behavior at high temperatures
over an unexpectedly broad doping range.
This indicates that
the high energy spin waves may survive even substantial doping,
which would have important implications for superconductivity.

In view of the above discussion, we present a study of
the quantum-critical behavior in the doped 2D t-J model using the
high temperature series expansion approach \cite{Singh:Glenister}.
The nearest-neighbor
version of the t-J model is described by the Hamiltonian:
\begin{equation}
\hat H = - t \sum_{\langle ij\rangle}
P (c^\dagger _{i\sigma} c_{j\sigma}+ \mbox{h.c.}) P +
J \sum_{\langle ij\rangle} {\bf S}_i {\bf S}_j.
\end{equation}
($P$ is the projection operator prohibiting double occupancy).
The 10-term
series in $\beta\equiv 1/T$ for the equal time correlation function,
$S({\bf q})$, and static susceptibility, $\chi({\bf q})$,
has been generated for arbitrary $t/J$,
number of electrons per unit cell $\rho=1\!-\!x$,
as well as for arbitrary ${\bf q}$.
The substitution of the expansion variable
$w \!=\! \tanh (\beta/\beta_0) $, which eliminates the influence
of any singularities outside the stripe
$|\mbox{Im} \beta | < \pi\beta_0/2 $,
has been applied in order to improve the convergence; several different
values of $\beta_0$
have been used and the consistency of the approximations have
been checked. Although our qualitative results
hold for a broad range of $t/J$ ratio,
we chose to present our data for one particular value of $t/J$=1 because
the accuracy improves for smaller $t/J$.

Our first important result is that in a broad range of the doping $x$,
temperature, and $t/J$ ratio,
$S({\bf q})$ and $\chi({\bf q})$ near ${\bf Q}=(\pi,\pi)$
are well described by the following scaling expressions:
\begin{equation}
S({\bf q}) = S_Q \cdot \hat S_m(\widetilde q \xi_m), \ \ \
\chi({\bf q}) = \chi_Q \cdot \hat \chi_m(\widetilde q \xi'_m),
\label{static:scaling}
\end{equation}
where ${\bf \widetilde q}=(q_x\!-\!\pi,q_y\!-\!\pi)$,
$ \hat S_m $ and $ \hat \chi_m $ are {\em universal}
scaling functions, while
$ S_Q, \ \chi_Q, \ \xi_m, \ \xi'_m $ depend on T, $x$, and $t/J$.
We define $\hat S_m(\kappa)$ and $\hat \chi_m(\kappa)$ so that
$ \hat S_m(0)=\hat \chi_m(0) = 1$
and $ d^2 \hat S_m/d\kappa^2|_{\kappa=0} =
d^2 \hat \chi_m /d\kappa^2|_{\kappa=0} = -2 $.
In this case, $\xi_m$ and $\xi_m'$ correspond to the ``second moment''
definition of the correlation length for $S({\bf q})$ and $\chi({\bf q})$,
respectively. Gradual character of deviation from the universal
behavior described by Eq.(\ref{static:scaling})
does not allow us to define an unambiguous
boundary at which Eq.(\ref{static:scaling}) fails.
For $t/J$=1,
the approximate range where the scaling behavior holds well
is T=0.6J-J, $x\!=\!0\!-\!15\%$;
more detailed discussion will be given in a
subsequent publication \cite{Glenister:Singh:Sokol}.

The scaling functions
$\hat S_m$ and $\hat \chi_m$ are computable in the QNL$\sigma$ model.
We observe that the 1/N calculation of Ref.\cite{Chubukov:Sachdev}
indicates negligible 1/N corrections to $\hat \chi_m $ and
$ \hat S_m $ in the QC regime
(note that 1/N corrections to the prefactors and $\xi$
are not negligible).
Therefore, we can safely use scaling functions calculated for
$N=\infty$, which we plot in Fig.\ref{profiles} along with our data.
We find that calculated $\hat S_m$ and $\hat \chi_m$
are very close to those obtained
from the collapsed data for the t-J model
(Fig.\ref{profiles}).
Further, in the asymptotic QC regime,
1/N calculation predicts
that $\hat \chi(\kappa)_m$
is nearly Lorentzian, while $\hat S(\kappa)_m$ larger than
Lorentzian for any given $\kappa$.
On the other hand, in the asymptotic RC regime,
$\hat S(\kappa)_m$ and $\hat \chi(\kappa)_m$ coincide and both
are smaller than Lorentzian.
We therefore conclude that the scaling functions
of Eq.(\ref{static:scaling}) indicate quantum critical behavior
in the t-J model at moderate temperatures.

We now turn to the dynamical properties. Since
\begin{eqnarray}
S({\bf q}) &=& \frac{1}{\pi}
\int d\omega \ g(\omega/T) \ \frac{\chi''({\bf q},\omega)}{\omega},
\nonumber \\
\chi({\bf q}) &=& \frac{1}{\pi}
\int d\omega \ \frac{\chi''({\bf q},\omega)}{\omega},
\end{eqnarray}
where $g(\zeta) = \frac{1}{2} \zeta /\tanh \left(\frac{1}{2}\zeta\right)$,
the ratio $S({\bf q})/T\chi({\bf q})$
reflects the frequency
distribution of the spectral weight at ${\bf q}$.
We find that for zero doping this ratio at ${\bf q}\!=\!(\pi,\pi)$
varies less than 1\% in the
range T=0.6J-J and is in excellent agreement with
the O(N) sigma model calculation of
Ref.\cite{Chubukov:Sachdev}:
\begin{equation}
\frac{S_Q}{T\chi_Q} \simeq
\left\{
\begin{array}{l}
1.09 \pm 0.01 \mbox{ (t-J)}, \\
1.09 \mbox{ (N=3) and } 1.08 \mbox{ (N=$\infty$)};
\end{array}
\right.
\end{equation}
in the asymptotic RC regime, this ratio would be equal to unity.
In the doped case, the ratio somewhat increases, up to $1.16$ for
$15\%$ doping at $T=J$. This increase may be caused
by a broad electron-hole continuum, which does not significantly modify
$\chi''({\bf q},\omega)$ for small frequencies, but can
yield significant contribution to the ratio because
$g(\zeta)$ is large for $\zeta\gg 1$.

Another universal quantity which may be temperature independent only
when $\bar{\omega}\sim T$, i.e.\ in case of QC behavior,
is $1-\xi_m/\xi'_m$. For the insulator,
this quantity indeed has
less than 7\% variation in the range T=0.6J-J.
The comparison to $O(N)$ calculations of Ref.\cite{Chubukov:Sachdev}
for the asymptotic QC regime of the QNL$\sigma$ model
yields:
\begin{equation}
1-\frac{\xi_m}{\xi'_m} \simeq
\left\{
\begin{array}{l}
0.043 \pm 0.003 \mbox{ (t-J)}, \\
0.043 \mbox{ (N=3) and } 0.035 \mbox{ (N=$\infty$)}.
\end{array}
\right.
\end{equation}
In the doped case, $\xi_m/\xi'_m$
becomes 10\% smaller at T=0.6J, a decrease that may
also be caused by the electron-hole continuum; it remains
nearly doping independent at T=J.

On the basis of the above arguments, we conclude that
the dynamical susceptibility of the t-J model can be written as:
\begin{equation}
\chi({\bf q},\omega) = \chi_Q \cdot \hat
\chi \left(\widetilde q \xi, \frac{\omega}{\bar \omega} \right)
\label{scaling}
\end{equation}
where to the leading order $\bar \omega \sim T$.
In addition,
in the asymptotic QC regime of the QNL$\sigma$ model,
not only Eq.(\ref{scaling}) holds, but it is also expected that
$ \chi_Q \sim \xi^{2-\eta} $,
where the critical exponent $\eta$
is nearly zero and can be neglected,
and $ 1/\xi\!\sim\!T $.
We first calculate $ \chi_Q/\xi^2 $ by generating 9-term series
directly for this quantity and find that for T=0.6J-J it
varies not more than 16\%, compared to far greater
(nearly by the factor of five in the undoped case) change in
$\chi_Q$ and $\xi^2$ separately.

We now turn to the temperature dependence of
$\xi$ and evaluate it by generating 9-term series for $\xi T^{1/2}$
and 10-term series for $ S_Q $ and $ \chi_Q T $.
$\xi^{-1}(T,x)$ for $t/J\!=\!1$ is plotted
as a function of temperature in Fig.\ref{xi1}.
As one can see, in a broad range of doping and temperature
$\xi^{-1}$ is nearly linear in $T$ with doping independent
slope.
As shown by Chakravarty \cite{Chakravarty},
one would expect the temperature dependence to be of the
form:
\begin{equation}
\frac{1}{\xi} \sim \frac{bT}{c} - \rm  C(x) \, T^{1-1/\nu}
\label{xi:universal}
\end{equation}
with the critical exponent $\nu \sim 0.7$. The second term varies slowly as
a function of temperature over the interval of comparisons,
and provides a doping dependent intercept for nearly linear
$\xi^{-1}(T)$. We note, however, that the
slope of $\xi^{-1}(T)$ determined from the numerical data
differs nearly by the factor of two from the value calculated
using Eq.(\ref{xi:universal}) with T=0 value
of $c$ and $b\simeq 0.8-1$ determined in Refs.
\cite{CHN,Chubukov:Sachdev,Manousakis:Salvador}.
We conjecture that the difference
is caused by the lattice corrections above T=0.6J.
Indeed, $\xi$ for the Heisenberg model is quite close to the value given
by Eq.(\ref{xi:universal}) at T=0.6J \cite{Chubukov:Sachdev}
and the deviation occurs only at higher temperatures.
On the other hand, our data unambiguously shows
that such corrections do not modify universal scaling functions of
Eqs.(\ref{static:scaling},\ref{scaling}). Note that
were the spin waves absorbed by the electron-hole background
in this doping range, $1/\xi^2$ and not $1/\xi$
would be linear in temperature \cite{Millis:z=2,SCR}.
Further, in the t-J model $1/\xi^2$ and not $1/\xi$ is linear
in temperature for $T\!\gg\!J$.
The measurements of Ref.\cite{Keimer} also indicate
linear $\xi^{-1}(T,x)$ with nearly doping independent slope.

Once it is established that the spin dynamics of doped
antiferromagnets is described by Eq.(\ref{scaling}),
we can now address the experimental result of
Imai, Slichter, and collaborators \cite{ImaiT1} that
at high temperatures the spin-lattice relaxation rate, $1/T_1$,
is nearly doping and temperature independent
in the doping range 0-15\%. From Eq.(\ref{scaling}) one obtains:
$ 1/T_1 \sim \smallint d{\bf q} \lim_{\omega\to 0}
\chi''({\bf q},\omega)/\omega \sim \chi_Q/\xi^2 $.
Indeed, we find that $\chi_Q /\xi^2 $ and therefore $1/T_1$
is nearly doping and temperature independent in a broad doping range.
Earlier, it has been shown \cite{Chubukov:Sachdev} that near the
critical point, temperature and parameter independence of
$1/T_1$ follows from the Josephson hyperscaling law;
our results show that such arguments are applicable even for
substantial doping.

Now we turn to the Gaussian component
of the spin-echo decay rate, $1/T_{2G}$, which is given by
(after Pennington and Slichter, \cite{Pennington:Slichter}):
\begin{equation}
\frac{1}{T_{\rm 2G}^2}\!=\!\frac{0.69}{8} \sum_{r\neq 0} a^2_r,
\ \
a_r\!=\!-\int \frac{d^2{\bf q}}{(2\pi)^2}
e^{i{\bf qr}} F_q^2 \, \frac{\chi({\bf q})}{g^2\mu_B^2},
\end{equation}
where $\chi({\bf q})$ is the static susceptibility and $F_q$ hyperfine
formfactor.
We evaluate this quantity by generating series directly
for $T/T_{\rm 2G}$ and then using Pade approximants.
We take $J=1500$K (in the region of experimental comparisons,
calculated $1/T_{\rm 2G}$ is not sensitive to the choice of $J$) and
the values of hyperfine couplings
determined for YBa$_2$Cu$_3$O$_{6.63}$ in
Ref.\cite{Monien:Pines:Takigawa}. The motivation for doing so is that
a number of experiments \cite{Imai:Slichter:pc}
indicate that the hyperfine couplings
do not change substantially from $\rm La_2CuO_4$
to $\rm YBa_2Cu_3O_{6+x}$.
Our results for $1/T_{\rm 2G}$
are plotted in Fig.\ref{T2G} along with the experimental data
of Imai, Slichter, and collaborators
\cite{ImaiT2} for the insulator. The results are
in excellent agreement with the experiments;
note that no adjustable
parameters are used.
We also find good agreement at high temperatures ($T\!>\!J/2$)
with the $4\!\times\!4$ cluster calculation for $x\!=\!0$
\cite{Sokol:Gagliano:Bacci}.

Recently, linear in temperature $T_{\rm 2G}$ above 200K has been
reported by Takigawa \cite{TakigawaT2}, which is consistent with the
QC prediction in the range of comparisons.
The slope of the linear high temperature part of the data
\cite{TakigawaT2} is larger in YBa$_2$Cu$_3$O$_{6.63}$ than in
La$_2$CuO$_4$;
in our study, we indeed find that the slope increases with increasing
doping and $t/J$ ratio (Fig.\ref{T2G}, inset).

To summarize, in our high temperature
series expansion study we find that
spin fluctuations in the t-J model exhibit quantum critical scaling
behavior. Particularly, for a broad range
of $t/J$ ratio, doping, and temperatures (1) the numerical data for
both $S({\bf q})$ and $\chi({\bf q})$ collapse to the universal
scaling function computable in the O(N) QNL$\sigma$ model,
(2) the characteristic energy scale for spin fluctuations,
$\bar{\omega}$, is proportional to temperature and
(3) $1/\xi$ is nearly linear in temperature. The disagreement
of the slope of $1/\xi$ with the prediction based on fully renormalized
T=0 value of $c$ is likely to be caused by
the lattice corrections above T=0.6J. One may
speculate that the absence of such corrections to the scaling functions
of Eqs.(\ref{static:scaling},\ref{scaling})
is related to the fact that in the QNL$\sigma$ model
they are much less dependent on the cutoff and N
(for a relevant discussion see Ref.\cite{Sachdev:unpublished}).
Our results are consistent with the conclusion of
Refs.\cite{Sachdev:Ye,Chubukov:Sachdev} that in the Shraiman-Siggia
model \cite{Shraiman:Siggia}
the presence of fermions does not alter the universality class
of the $z\!=\!1$ critical point separating phases with Neel and
short range order.
While the t-J model might not be quantitatively
applicable for the doped cuprates, we believe
that a broad region of the quantum critical behavior
is a general feature of moderately doped antiferromagnets.

In the second part of our study, we calculated the spin-echo decay
rate, $1/T_{\rm 2G}$, recently measured in a number of the high-T$_c$
compounds.
Our results for $x\!=\!0$ (i.e.\ in the Heisenberg model),
obtained with no adjustable parameters
are in excellent agreement
with the experiments in La$_2$CuO$_4$ \cite{ImaiT1,ImaiT2}
and show that the hyperfine couplings do not vary
substantially from YBa$_2$Cu$_3$O$_{6.63}$ to La$_2$CuO$_4$.
We also find qualitative agreement with the experimental data on
$1/T_{\rm 2G}$ in YBa$_2$Cu$_3$O$_{6.63}$ \cite{TakigawaT2} and
show that for moderate doping,
$1/T_1$ is nearly doping and temperature independent,
in agreement with both the experimental data in $\rm La_{2-x}Sr_xCuO_4$
\cite{ImaiT1} and earlier analysis
based on the QNL$\sigma$ model \cite{Chubukov:Sachdev}.

Our results provide additional support to
the conjecture of Ref.\cite{Sokol:Pines} that the
high temperature quantum critical
behavior \cite{CHN,Sachdev:Ye,Chubukov:Sachdev} is a common feature
shared by many of the cuprate superconductors.
More detailed study of
this and related subjects will be presented in a subsequent publication
\cite{Glenister:Singh:Sokol}.

We are grateful to T. Imai, C.P. Slichter,
and M. Takigawa for communicating their experimental results to us
prior to publication, and to R.J. Birgeneau,
S. Chakravarty, A.V. Chubukov, E. Dagotto, M.P. Gelfand,
L.P. Gor'kov, T. Imai, D. Pines, W.O. Putikka, S. Sachdev
and C.P. Slichter for discussions.
This work has been supported by the
NSF Grants DMR89-20538 through the Materials Research Laboratory
of the University of Illinois at Urbana-Champaign and DMR90-17361.

\begin{figure}
\caption{}
Universal scaling functions
$\hat \chi(\kappa)$ ({\protect\large$\bullet$} and solid line)
and $\hat S(\kappa)$ ({\protect\large$\circ$} and dashed line).
Symbols are obtained by collapsing our numerical data for the t-J model.
Lines show analytical predictions for the
O(N) QNL$\sigma$ model;
results of Ref.\cite{Chubukov:Sachdev} indicate that there is
virtually no dependence on N for $N\ge3$. Note that the lines are
universal scaling predictions rather than fits.
\label{profiles}
\end{figure}

\begin{figure}
\caption{}
$\xi^{-1}(T,x)$ for $t/J$=1 and
({\protect\large$\circ$})
the experimental
data of Ref.\protect\cite{Keimer} for the undoped La$_2$CuO$_4$
(we set $a$=1);
our calculation
falls within not shown experimental errorbars.
\label{xi1}
\end{figure}

\begin{figure}
\caption{}
The spin-echo decay rate $1/T_{\rm 2G}(T)$
in the Heisenberg model (i.e.\ for x=0) calculated without
adjustable parameters and
({\protect\large$\bullet$})
the experimental data of
Ref.\protect\cite{ImaiT2} for La$_2$CuO$_4$.
Inset: $T_{\rm 2G}(x,T)$ for $t/J$=1.
\label{T2G}
\end{figure}

\end{document}